\documentclass[twocolumn]{svjour3}          
\smartqed  
\usepackage{graphicx}

\usepackage{booktabs}
\usepackage{pgf}
\usepackage{pgfplots,filecontents,pgfplotstable}
\usepackage{graphicx}  
\usepackage{epstopdf} 
\usepackage{tikz}
\usetikzlibrary{pgfplots.groupplots}

\journalname{Computational Astrophysics and Cosmology}
\begin{document}

\title{PKDGRAV3: Beyond Trillion Particle Cosmological Simulations
for the Next Era of Galaxy Surveys}


\author{Douglas Potter  \and
        Joachim Stadel  \and
        Romain Teyssier
}


\institute{Douglas~Potter \and Joachim~Stadel \and Romain~Teyssier \at
  douglas.potter, joachim.stadel, romain.teyssier @uzh.ch \\
              University of Zurich, Zurich, Switzerland\\
}

\date{Received: date / Accepted: date}

\maketitle

\begin{abstract}
We report on the successful completion of a 2 trillion particle
cosmological simulation to z=0 run on the Piz Daint supercomputer 
(CSCS, Switzerland), using 4000+ GPU nodes for a little less than 80h 
of wall-clock time or 350,000 node hours.
Using multiple benchmarks and performance measurements on the US
Oak Ridge National Laboratory Titan supercomputer, we demonstrate
that our code PKDGRAV3, delivers, to our knowledge,
the fastest time-to-solution for large-scale cosmological N-body simulations.
This was made possible by using
the Fast Multipole Method in conjunction with individual and adaptive
particle time steps, both deployed efficiently (and for the first time)
on supercomputers with GPU-accelerated nodes. 
The very low memory footprint of PKDGRAV3 allowed
us to run the first ever benchmark with 8 trillion particles 
on Titan, and to achieve perfect scaling up to 18000 nodes 
and a peak performance of 10 Pflops.
\end{abstract}


\section{Overview of the Problem}
\label{overview}
The last decade has seen the advent of high precision cosmology, mostly because of  the very accurate Cosmic Microwave 
Background (CMB) experiments WMAP \cite{2003ApJS..148..175S} and Planck \cite{Ade:2014dt}.
Cosmological parameters, such as the total matter content in the Universe or the Hubble constant are now 
constrained to within several percent. 
Although our best fit model, the so-called standard Lambda Cold Dark Matter (LCDM) model, very successfully explains
these remarkable observations, it is still based on two mysterious, undetected and elusive components: 
dark matter and dark energy. The  cosmological experiments  of the next decade might
shed light on this ``dark sector'' and possibly revolutionize modern physics. 
After a decade of CMB experiments, we expect large scale galaxy surveys, 
such as the ground based Large Synoptic Survey Telescope \cite{Collaboration:2009vz} (LSST), 
or the two satellite missions Euclid \cite{2011arXiv1110.3193L} (in Europe) 
and WFIRST \cite{2013arXiv1305.5422S} (in the US), to give new, stronger constraints on our 
standard cosmological model parameters, possibly below the percent level. 
Two techniques are considered to measure the clustering of matter as a function of time 
and scale: weak lensing (WL) and galaxy clustering (GC). Both techniques rely on very accurate 
theoretical predictions of the non-linear dynamics of the dark matter fluid in an expanding Universe. 
The more accurate these theoretical predictions are,
the more efficient the future large scale surveys will be in solving the mysteries of the dark universe. 

\begin{figure}
\centering
\includegraphics[width=\columnwidth]{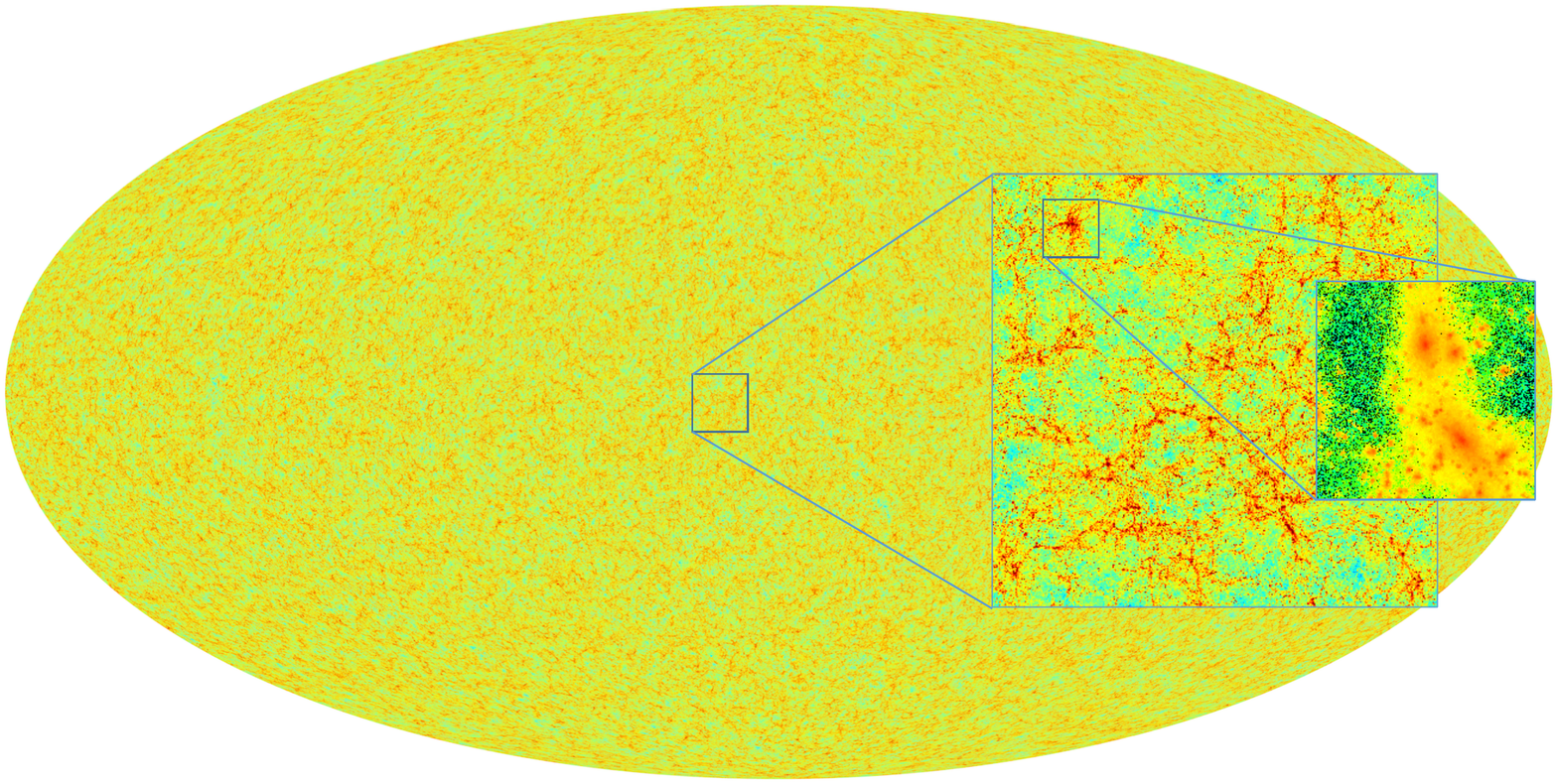}
\caption{Simulated full-sky matter distribution from a 2 trillion particles simulation. The zoom-in
quadrant shows the non-linear, filamentary structure of the universe on small scale. }
\label{fig:fullsky}
\end{figure}

Because of the non-linear nature of gravity on these scales, our best theoretical predictions make use of $N$-body 
simulations: the dark matter fluid is sampled in phase space using as many macro-particles as possible, each one 
representing a large ensemble of true, microscopic dark matter particles, evolving without collision under the 
effect of their mutual gravitational attraction. We review in Section~\ref{stateofart}
the current state of the art in the development of high performance $N$-body codes. 
Motivated by future dark energy missions, our main goal is to reach an accuracy better than 1\%
in the power spectrum of the matter density field from linear scales ($>100$~Mpc/h) 
down to strongly non-linear scales ($\simeq 1$~Mpc/h).
For us to reach these extreme accuracy requirements, we face four different computational challenges: 
1- high precision in the gravity calculation, 
2- high accuracy in the time stepping, 
3- reduce the statistical errors below 1\%, which translates to a physical
volume of $L \simeq 2$~Gpc/h, and 
4- high enough mass resolution, that translates to a large number of particles (for a review see 
Ref.~\cite{2016JCAP...04..047S}). 
The last requirement pushes the limits of what can be achieved on current supercomputers: 
we need to model accurately dark matter haloes as small as one tenth of the Milky Way mass, 
which translates into a particle mass smaller than $10^9$~M$_\odot$/h, and, for the adopted minimum 
box size, into a total particle count of $N$ $>$ 2 trillion. In the context of future large galaxy 
surveys, we will need these extreme $N$-body simulations not just once, but for many different cosmological 
models, exploring alternative gravity models or galaxy formation scenarios. An additional requirement is 
a fast enough time-to-solution, so that $N$-body simulation can optimize and analyze cosmology experiments. 

In this paper, we report on the successful evolution of a 2 trillion particles 
simulation of the LCDM model from $z=49$ to $z=0$
{\it in less than 80h of wall clock time} including on-the-fly analysis,
performed on the  the Swiss National Supercomputing Center Machine, Piz Daint, 
using $4000+$ GPU-accelerated nodes.
We also report on the first ever benchmark of a 8 trillion 
particles simulation of the same model, performed on Titan
at Oak Ridge using 18000 GPU-accelerated nodes. 
Although our 2 trillion particles run represents the minimum 
requirements for future galaxy surveys, we 
establish the feasibility of even more extreme particle counts 
with our 8 trillion particle benchmark. Our tests demonstrate a significant 
reduction in the time-to-solution 
and put us in an ideal position to use these extreme $N$-body 
simulations for the preparation and the analysis of large 
galaxy surveys. 

\section{Current State of the Art}
\label{stateofart}

$N$-body simulations in astrophysics 
have been at the forefront of high performance computing,
even before the first digital computer, with the galaxy collision
experiment of 
Holmberg \cite{1941ApJ....94..385H}, based on moving light bulbs, 
and then the heroic 300-particle computer simulation of the 
Coma cluster performed by Peebles in 1969 \cite{1970AJ.....75...13P}.
Cosmolo\-gi\-cal simulations have been particularly 
efficient at exploiting the best of each generation of supercomputers, 
adapting the algorithms to new architectures. 
In that respect, the number 
of simulated bodies (or particles) has increased dramatically, owing 
to the ever increasing performance of supercomputers, but also to the 
growing efficiency of the $N$-body solvers. Here, we report
the first benchmark ever performed on 8 trillion 
($8\times 10^{12}$) particles.

In the early 80's, gravity calculations quickly moved away from the accurate but slow 
$\mathcal{O}(N^2)$ direct interaction (where $N$ stands here for the number of 
simulated particles) or Par\-ticle-Particle (PP) approach, to faster
techniques, such as the Particle-Mesh (PM) 
algorithm \cite{1988csup.book.....H},
based on the Fast Fou\-rier Transform (with $\mathcal{O}(N \ln N)$ efficiency) 
or the tree code \cite{1986Natur.324..446B} (also with 
$\mathcal{O}(N \ln N)$ scaling).
Since the PM technique suffers from the limited resolution of the mesh,
a hybrid version of PP and PM was later developed, leading to the P$^3$M
technique, which is $\mathcal{O}(N \ln N)$ on large scale and 
$\mathcal{O}(N^2)$ on small scale \cite{1995ApJ...452..797C}.
The attitude of many generations of code developers since then
was to take advantage of the shear performance of
the best available computer at that time, but also to reduce drastically
the time-to-solution by developing more complex but more
efficient algorithms.

In that respect, cosmological simulations are particularly challenging,
since they require a fixed simulation time of 13.7~Gyr, 
namely from the Big Bang 
until our present epoch.
They also require, as explained in Section~\ref{overview}, the largest 
possible number of particles that can fit in the computer memory. This 
has led computational cosmologists to develop clever and innovative solutions
to optimize the gravity solvers. 

\pgfplotstableread{
0.00396493 5.82417 217.421 523.248 882.986 1710.97 3606.68 0.430674
0.00565417 10.8148 403.102 968.079 1631.75 3157.96 6667.98 0.302287
0.00733457 10.3105 388.424 937.181 1583.18 3071 6502.06 0.233643
0.00906711 12.9176 484.511 1165.87 1966.85 3809.6 8057.51 0.186133
0.0106963 14.8243 555.132 1335.11 2251.97 4361.48 9223.57 0.16271
0.0123615 13.9 520.666 1252.49 2112.56 4090.51 8645.61 0.139261
0.0140367 15.502 581.173 1398.68 2359.71 4570.3 9659.51 0.122275
0.0157228 14.855 556.735 1339.34 2258.99 4373.52 9240.13 0.108809
0.0173903 15.3753 576.196 1386.06 2337.75 4525.92 9561.67 0.0989224
0.0190316 14.6179 547.985 1318.39 2223.8 4305.68 9097.87 0.0911098
0.0207 15.5125 581.204 1398.09 2357.77 4563.19 9632.79 0.0830834
0.0224027 15.1831 568.748 1367.85 2306.52 4463.43 9420.82 0.0759915
0.0240623 14.1753 531.113 1277.52 2154.3 4168.73 8796.9 0.0716631
0.0257122 14.3398 537.133 1291.67 2177.65 4212.18 8881.82 0.0672615
0.0273696 13.5425 507.332 1220.46 2058.44 3984.61 8412.79 0.0630453
0.0290314 13.1865 493.806 1187.42 2001.97 3872.77 8167.59 0.0593577
0.0306995 12.2075 457.302 1099.72 1854.01 3585.76 7558.51 0.0560263
0.0332408 11.7596 440.316 1058.63 1784.58 3451.25 7274.42 0.0419213
0.035699 11.1327 416.986 1002.72 1690.48 3269.48 6890.84 0.039689
0.0382282 10.1669 381.02 916.453 1545.28 2989.27 6301.89 0.0365392
0.0406805 9.71371 363.728 874.587 1474.44 2851.59 6009.96 0.0348708
0.043202 9.11777 341.75 822.029 1386.08 2681.26 5652.06 0.0323818
0.046525 8.44488 316.269 760.517 1282.12 2479.35 5223.08 0.0261929
0.0498347 8.04884 301.553 725.216 1222.64 2364.26 4979.76 0.0245024
0.0531681 7.70162 288.64 694.344 1170.76 2264.21 4768.51 0.0228844
0.0565101 7.30991 274.036 659.316 1111.87 2150.96 4532.48 0.0215033
0.0607022 7.00148 262.279 630.701 1063.17 2055.25 4325.18 0.0178737
0.0656758 6.56653 245.865 591.039 996.067 1924.7 4047.51 0.0151668
0.0697964 6.27949 235.357 566.115 954.452 1845.35 3883.22 0.0156791
0.073996 5.87393 219.959 528.796 891.167 1721.83 3619.27 0.0146495
0.0798564 5.3879 201.819 485.314 818.03 1580.9 3324.05 0.0114911
0.0856279 4.79074 179.455 431.588 727.648 1407.1 2962.82 0.0107988
0.0914779 4.25478 159.393 383.46 646.766 1251.87 2641.5 0.0100402
0.0981251 3.76264 141.061 339.607 573.235 1111.24 2351.84 0.00878082
0.104763 3.37239 126.432 304.489 514.172 997.724 2116.18 0.00823022
0.112276 3.08861 115.861 279.176 471.658 916.016 1945.58 0.0072184
0.120587 2.88314 108.203 260.806 440.726 856.197 1818.81 0.0063901
0.12889 2.69466 101.112 243.702 411.807 799.952 1698.98 0.00598133
0.137194 2.48428 93.1992 224.632 379.613 737.588 1567.54 0.00561896
0.147241 2.1753 81.6889 197.163 333.704 650.485 1391.22 0.00475979
0.158 1.8555 69.8082 168.898 286.618 561.755 1214.14 0.00428639
0.168834 1.63521 61.5978 149.292 253.832 499.479 1087.74 0.00399742
0.180463 1.4887 56.0983 136.028 231.401 455.813 994.639 0.00360973
0.192955 1.37302 51.7462 125.483 213.457 420.383 916.942 0.00325734
0.207108 1.21593 45.8794 111.453 189.957 375.57 825.057 0.00285111
0.222059 1.03888 39.2998 95.8181 163.966 326.857 728.845 0.00258721
0.237011 0.914912 34.702 84.8708 145.695 292.215 658.436 0.00242391
0.253695 0.823044 31.2319 76.4543 131.386 264.11 597.895 0.00214374
0.271963 0.728062 27.6697 67.8849 116.941 236.229 539.687 0.00191108
0.291104 0.628478 23.9804 59.1324 102.399 208.909 484.836 0.00174423
0.311908 0.550203 21.0453 52.0773 90.5201 186.034 437.195 0.00156147
0.333501 0.48753 18.6807 46.343 80.7748 166.939 396.47 0.00143344
0.356829 0.422511 16.2588 40.5571 71.0885 148.461 358.341 0.00128895
0.381744 0.369084 14.2504 35.71 62.8931 132.553 324.746 0.00116582
0.408392 0.322582 12.5 31.4751 55.7104 118.518 294.859 0.00105372
0.437498 0.279127 10.8708 27.5513 49.0883 105.691 267.647 0.000941168
0.468264 0.242626 9.48708 24.1837 43.3476 94.3829 243.292 0.00085528
0.501558 0.209708 8.24676 21.1757 38.2352 84.3484 221.582 0.000767592
0.537308 0.181229 7.16588 18.5356 33.7187 75.398 201.973 0.00069147
0.575585 0.156227 6.21781 16.2184 29.748 67.492 184.559 0.000623815
0.616314 0.134719 5.39796 14.2044 26.2807 60.5283 168.972 0.000564782
0.659573 0.116183 4.68762 12.4498 23.2484 54.407 155.149 0.000512075
0.706155 0.0999131 4.06332 10.9028 20.5618 48.9286 142.555 0.00046092
0.75606 0.0858537 3.52008 9.54818 18.1996 44.0883 131.299 0.000415917
0.809285 0.073708 3.049 8.36644 16.1257 39.7958 121.165 0.000376249
0.866701 0.063162 2.63828 7.33051 14.2959 35.9472 111.841 0.000338259
0.928231 0.0540533 2.28095 6.42214 12.6826 32.5281 103.451 0.000305095
0.993082 0.0463261 1.97644 5.64305 11.2887 29.5364 95.8821 0.000277774
1.06299 0.0396018 1.70941 4.95537 10.0523 26.8553 88.9351 0.000249944
1.13868 0.033768 1.47574 4.34731 8.94919 24.427 82.4778 0.00022424
1.21935 0.0287913 1.27496 3.82028 7.9854 22.2749 76.5637 0.000202838
1.305 0.0245635 1.10272 3.36365 7.14238 20.3616 71.1175 0.000183933
1.39737 0.0209089 0.952688 2.96169 6.3932 18.6313 66.0295 0.000165408
1.49631 0.0177794 0.822893 2.61037 5.7327 17.0748 61.2643 0.000149254
1.60197 0.0151128 0.711081 2.30423 5.15093 15.6747 56.8014 0.000134904
1.71506 0.0128384 0.614639 2.0365 4.63612 14.4095 52.5975 0.000121799
1.83654 0.0108936 0.531061 1.80158 4.17914 13.257 48.6207 0.000109744
1.96624 0.00923817 0.459152 1.59652 3.77514 12.2106 44.8657 9.92037e-05
2.10519 0.00782724 0.397031 1.4167 3.41573 11.2524 41.2998 8.95187e-05
2.25403 0.00662716 0.343392 1.25907 3.09597 10.3739 37.91 8.07821e-05
2.41292 0.00561132 0.297251 1.12117 2.81181 9.56524 34.7081 7.30371e-05
2.5843 0.00474305 0.257175 0.999106 2.55606 8.81163 31.6538 6.56616e-05
2.76722 0.00400694 0.222649 0.892197 2.32766 8.11383 28.7764 5.93554e-05
2.96272 0.00338397 0.192932 0.798221 2.12307 7.46478 26.0742 5.36254e-05
3.1723 0.00285588 0.16724 0.715252 1.93843 6.85698 23.5376 4.8371e-05
3.39604 0.00241093 0.145116 0.642214 1.77211 6.28932 21.1809 4.3731e-05
3.63646 0.00203292 0.125944 0.577373 1.62079 5.75437 18.9865 3.93977e-05
3.89428 0.00171233 0.109345 0.519862 1.48283 5.25146 16.9581 3.55262e-05
4.16958 0.00144223 0.0950467 0.468921 1.35724 4.7806 15.1037 3.21099e-05
4.46402 0.00121479 0.0826961 0.423655 1.24228 4.33934 13.4131 2.90008e-05
4.77926 0.00102315 0.0720182 0.383262 1.13655 3.92635 11.8787 2.6179e-05
5.11782 0.000861083 0.0627557 0.347037 1.03873 3.53999 10.4895 2.35902e-05
5.48043 0.000724333 0.0547467 0.314581 0.948291 3.1806 9.24098 2.12863e-05
5.86803 0.000609529 0.0478365 0.285483 0.864692 2.84868 8.12651 1.92287e-05
6.28308 0.000513242 0.0418678 0.259292 0.787176 2.54313 7.13399 1.73545e-05
6.72723 0.000432659 0.0367208 0.235728 0.715396 2.26369 6.25376 1.56687e-05
7.20301 0.000365195 0.0322915 0.214563 0.649172 2.0098 5.47614 1.4139e-05
7.71201 0.000308933 0.0285086 0.19572 0.588607 1.78159 4.79409 1.27676e-05
8.25683 0.000262402 0.0253282 0.179293 0.534292 1.57975 4.20263 1.15264e-05
8.84072 0.000224392 0.0227634 0.165727 0.487688 1.40688 3.70277 1.03988e-05
9.46622 0.000194088 0.0209576 0.156315 0.452577 1.27191 3.31265 9.38305e-06
10.1358 0.000171344 0.0203455 0.154345 0.438361 1.19909 3.0908 8.46984e-06
}\pks

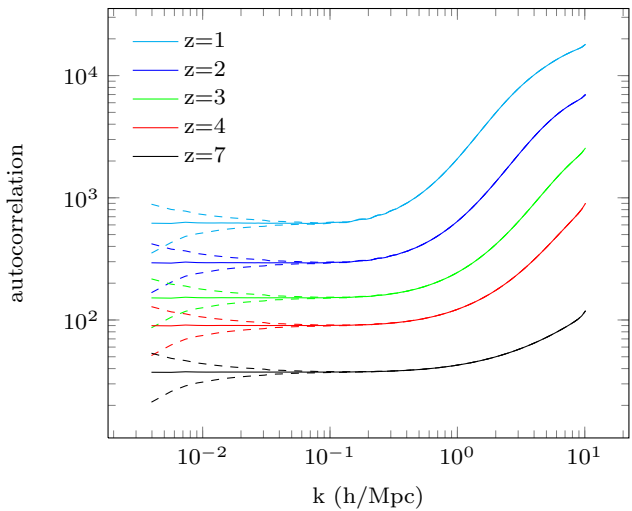
\begin{figure}
\centering
\begin{tikzpicture}[scale=1.0]
\begin{loglogaxis}[
  legend pos=north west,
  legend style={draw=none},
  xlabel={k (h/Mpc)},
  ylabel={autocorrelation}]
\addplot[cyan,no marks] table[x index=0,y expr=\thisrowno{6}/\thisrowno{1}]{\pks};
\addlegendentry{z=1};
\addplot[cyan,dashed,forget plot,no marks] table[x index=0,y expr=\thisrowno{6}/\thisrowno{1}+\thisrowno{7}*(\thisrowno{6}+(2000/12600)^3)/\thisrowno{1}]{\pks};
\addplot[cyan,dashed,forget plot,no marks] table[x index=0,y expr=\thisrowno{6}/\thisrowno{1}-\thisrowno{7}*(\thisrowno{6}+(2000/12600)^3)/\thisrowno{1}]{\pks};

\addplot[blue,no marks] table[x index=0,y expr=\thisrowno{5}/\thisrowno{1}]{\pks};
\addlegendentry{z=2};
\addplot[blue,dashed,forget plot,no marks] table[x index=0,y expr=\thisrowno{5}/\thisrowno{1}+\thisrowno{7}*(\thisrowno{5}+(2000/12600)^3)/\thisrowno{1}]{\pks};
\addplot[blue,dashed,forget plot,no marks] table[x index=0,y expr=\thisrowno{5}/\thisrowno{1}-\thisrowno{7}*(\thisrowno{5}+(2000/12600)^3)/\thisrowno{1}]{\pks};

\addplot[green,no marks] table[x index=0,y expr=\thisrowno{4}/\thisrowno{1}]{\pks};
\addlegendentry{z=3};
\addplot[green,dashed,forget plot,no marks] table[x index=0,y expr=\thisrowno{4}/\thisrowno{1}+\thisrowno{7}*(\thisrowno{4}+(2000/12600)^3)/\thisrowno{1}]{\pks};
\addplot[green,dashed,forget plot,no marks] table[x index=0,y expr=\thisrowno{4}/\thisrowno{1}-\thisrowno{7}*(\thisrowno{4}+(2000/12600)^3)/\thisrowno{1}]{\pks};

\addplot[red,no marks] table[x index=0,y expr=\thisrowno{3}/\thisrowno{1}]{\pks};
\addlegendentry{z=4};
\addplot[red,dashed,forget plot,no marks] table[x index=0,y expr=\thisrowno{3}/\thisrowno{1}+\thisrowno{7}*(\thisrowno{3}+(2000/12600)^3)/\thisrowno{1}]{\pks};
\addplot[red,dashed,forget plot,no marks] table[x index=0,y expr=\thisrowno{3}/\thisrowno{1}-\thisrowno{7}*(\thisrowno{3}+(2000/12600)^3)/\thisrowno{1}]{\pks};

\addplot[black,no marks] table[x index=0,y expr=\thisrowno{2}/\thisrowno{1}]{\pks};
\addlegendentry{z=7};
\addplot[black,dashed,forget plot,no marks] table[x index=0,y expr=\thisrowno{2}/\thisrowno{1}+\thisrowno{7}*(\thisrowno{2}+(2000/12600)^3)/\thisrowno{1}]{\pks};
\addplot[black,dashed,forget plot,no marks] table[x index=0,y expr=\thisrowno{2}/\thisrowno{1}-\thisrowno{7}*(\thisrowno{2}+(2000/12600)^3)/\thisrowno{1}]{\pks};

\end{loglogaxis}
\end{tikzpicture}
\caption{Auto-correlation functions of the density fluctuation in a cosmological box of 3~Gpc
sampled with 2 trillion particles simulation at various epoch indicated by the redshift. 
Dashed lines indicate the statistical errors due to the finite volume of the simulation. 
The accuracy of these theoretical predictions is far below the percent level on almost all scales.}
\label{fig:pk}
\end{figure}

Warren and Salmon were among the first cosmologists to be recognized
for their parallel tree code's performance, reaching 430 Gflops on 
ASCI Red \cite{Warren:1993tg,Warren:1997tu}.
In 2012, The Millennium XXL simulation\cite{2012MNRAS.426.2046A} was run
with 0.3 trillion particles using a specialized version of the GADGET-3 code,
based on GADGET-2\cite{2005MNRAS.364.1105S}.
At about the same time, Ishiyama {\it et al.} also achieved
4.5~Pflops with a 1 trillion particle simulation run on the 
K computer \cite{2012arXiv1211.4406I} for a cosmological simulation
using GreeM \cite{2009PASJ...61.1319I},
another parallel tree code. 
Habib {\it et al.} \cite{2013hpcn.confE...6H} performed
a $3.7\times 10^{12}$ particle benchmark on a BG/Q system in 2013, 
this time with a new generation
PM+X\footnote{HACC can use a number of hybrid PM methods including $P^3M$ or TreePM
with or without GPU or other accelerators.} code called HACC.
The HACC code was used in 2014 to produce
the Q Continnum Simulation\cite{2015ApJS..219...34H};
a full cosmological simulation of 0.55 trillion particles.
In 2014 another 1 trillion particle simulation was run
by Skillman {\it et al.} \cite{2014arXiv1407.2600S}
using the 2HOT code \cite{2013arXiv1310.4502W}.
More recently, Bedorf {\it et al.} \cite{2014hpcn.conf...54B}
developed a tree code
fully ported on GPUs, and delivered almost 
25~Pflops on the Titan supercomputer. 
These recent achievements
demonstrate that tree codes and P$^3$M codes, both scaling as 
$\mathcal{O}(N \ln N)$,
can deliver significant performance on parallel, and more recently on GPU 
accelerated, hardware. 

In parallel, however, new algorithms have been developed,
both for particle and grid-based gravity solver, which in principle 
could reduce even more the time-to-solution for cosmological simulations. 
These are the Multigrid (MG) solver \cite{1973LNP....19...82B},
which can replace the FFT advantageously, as it scales as $\mathcal{O}(N)$,
and the Fast Multipole Method (FMM) \cite{1987JCoPh..73..325G,2002JCoPh.179...27D}
which  could deliver the same $\mathcal{O}(N)$ scaling for tree-based codes.
While the former, implemented in the Adaptive Mesh Refinement code 
RAMSES \cite{2002A&A...385..337T}, has been used recently in the
500~billion particles
cosmological simulation DEUS \cite{2012arXiv1206.2838A}, the latter,
implemented in the PKDGRAV3 code, is the main subject of the present paper.

The $\mathcal{O}(N)$ scaling of FMM clearly offers the opportunity to 
go to higher particle counts, or to reduce significantly the time-to-solution
for a fixed $N$. Since cosmological simulations are targeting the highest 
possible value for $N$, memory is also a strong limitation. The
main innovations presented in this paper are  
1- a highly performing version of the FMM algorithm, with a measured peak performance of 10~Pflops,
and 2- an optimal use of the available memory, allowing us to reach 8 trillion particles on the 18000 nodes
of the Titan supercomputer.
\section{Algorithmic Improvements}
\subsection{Fast Multipole Method}
As the ``$N$'' in $N$-body simulations has increased into the trillions, 
the asymptotic order of the algorithms to calculate the gravitational forces between the particles
is central to having a fast time-to-solution. The $\mathcal{O}(N \ln N)$ gravity 
calculation of Barnes-Hut (BH) tree-codes, even high\-ly optimized ones which achieve 
excellent peak performance, are problematic for cosmology simulations.
FMM is now vastly superior to the BH for large $N$, even though 
it has somewhat lower peak floating point rate than measured by some recent BH codes 
(Bonzai\cite{2014hpcn.conf...54B}, 2HOT\cite{2013arXiv1310.4502W}).
An aspect of FMM for cosmology simulation is that unlike other codes (BH, 
P$^3$M, and tree-PM) the gravity calculation does not take longer as the simulation 
progresses from the early smooth state of the 
Universe toward the present day, highly clustered state of matter. This is because FMM
{\it must}, by its scaling with $N$, be effectively ``blind'' to the depth of the tree 
structure, and hence to the degree of clustering present among the particles in the 
simulation. FMM and BH are very similar methods; both use particle-particle (PP) interactions
for nearby particles and a multipole expansion of the mass within a more distant cell 
to approximate the force (PC-interactions). However, FMM also considers 
{\em cell-cell} (CC) interactions by approximating the potential ``landscape'' within a 
given cell (the sink cell) that is induced by a sufficiently distant multipole (the source cell). 
While any implementation which uses CC interactions in a sufficiently general way will scale as
$\mathcal{O}(N)$ and thus qualifies as an FMM code, several key differences make the FMM as used in 
PKDGRAV3 highly efficient for very large $N$ simulations.

FMM was originally implemented by Greengard \cite{1987JCoPh..73..325G} using a hierarchy of 
uniform meshes, but is in fact perfectly suited to implementation using a tree structure 
as in the BH method. Unlike most tree-codes, PKDGRAV3, uses a binary tree where parent cells 
are divided along the longest axis into two equal volumed child cells. Using a binary tree 
as opposed to an oct-tree provides a finer jump in accuracy when going from
an expansion based on a parent cell to using the sum of expansions for the child cells. 
This leads to fewer terms being required to achieve the same force calculation accuracy
at the expense of somewhat higher cost in making these decisions (tree walk phase).
Another advantage is the simplicity of handling the non-cubical domains that result from
{\em domain decomposition} which divides the simulation volume into sub-volumes which are local
to each core. Since we use the traditional ORB (Orthogonal Recursive Bisection) decomposition
to balance the number of particles in the domains, this forms the upper part of our global
tree structure of which each node and core has a purely local subtree. 
In fact FMM naturally maximizes locality even within the memory hierarchy as it 
proceeds down the tree toward the leaf cells since the particles and cells are in a
hierarchically sorted order after building the tree. Leaf cells of our tree contain up to
$b$ particles (we call this the {\em bucket size}), where the optimal value is around 16.

Central to the efficiency of a tree code, particularly one using GPU acceleration (see below), 
is how we create lists of interactions (PP, PC, CC and CP\footnote{Cell-particle interactions 
are the mirror image of particle-cell interactions; they are the expansions of the potential
within the sink cell induced by a single source particle.}) which when evaluated give us the force
on the particles. We walk the tree structure
in node-left-right recursive order for sink cells (to which interactions apply) considering
source cells that are collected on a checklist. Considering source cells for interactions
is traditionally referred to as evaluating an {\em opening criterion}, but opening a cell 
(removing it from the checklist and adding its children to the end of the checklist) is only
one possible outcome. A source cell on the checklist could also be put onto any of the 
four interaction lists depending on its distance from the sink cell, or it could remain on
the checklist for further consideration by the {\em children of the sink cell} as we proceed
deeper in the tree.\footnote{It is rare for a cell to stay on the checklist for more than a few 
levels as it will end up on one of the interaction lists or be opened.} 
Evaluating the opening criterion
is a purely arithmetic operation (using AVX/SSE intrinsics for performance and to avoid branches)
resulting in a case value of 1 to 6 encoding the outcome for checklist elements.
When done this way, these calculations are insignificant to the total computing cost ($\sim 2\%$).
Tree walking begins with the sink cell being the root of the {\em local} tree of a processor
while the checklist contains the {\em global} root cell of the entire simulation box as 
well as its 26 (and sometimes 124 depending on accuracy requirements) surrounding periodic 
replicas.


The actual opening criterion is critical in controlling the distributions of force errors, 
both in their magnitude and in their spatial correlations.\footnote{Ideally we want spatially 
uncorrelated errors, but this is as impossible to attain as is having all force errors
precisely at the desired truncation error.}
During tree build we calculate a bounding box for each cell and the distance, $b_{\rm max}$, 
from the center of mass of the cell (which is always the center of expansions in PKDGRAV3) 
to the most distant particle in the cell. Based on this we determine an {\em opening radius}
for a cell, $RO = b_{\rm max}/\theta$, where $\theta$ is the traditional opening angle 
and the force accuracy controlling parameter in the code. If the distance between the
source and sink (between centers of mass) are greater than $1.5 RO_{\rm sink}+RO_{\rm source}$ 
{\em and} the bounding boxes are no closer than twice the {\em softening} (we use
1/50 times the mean inter-particle separation -- for a review on the role of softening in 
$N$-body simulations see \cite{2011EPJP..126...55D}), then this is a CC or CP interaction. 
Note, that there is a deliberate asymmetry here, the factor of 1.5, which 
controls the spatial correlations in the force errors. For a traditional BH code the force
errors typically add up from all directions about a given particle and tend to be correlated 
spatially with the density of particles. For FMM on the other hand, there is almost no correlation 
with density (again a working FMM must be blind to tree depth), but we see the tree structure 
since the expansion of the potential within a sink cell is most accurate at the center of mass 
and degrades toward the edge of the cell. To reduce this spatial correlation below about 10\%
of the random errors we have made the acceptance of CC and CP interactions stricter by making 
sink opening radii larger by this factor.
If leaf cells are opened their particles are added to the checklist with 
$RO_{\rm source}=0$ and can later become CP or PP interactions. 
If a source cell is reached with fewer than $g$ particles (called the {\em group size}) we 
proceed no deeper in the tree resolving the remaining checklist into interaction lists, 
including now PP and PC as well.
We have found that a group size of 64, or more generally four times the bucket size, seems to
be close to optimal for PKDGRAV3.


Most tree-codes consider multipoles of up to only 2nd order (quadrupoles) which is 
most efficient for low accuracy force calculation, however for the needed force
accuracy of better than 0.1\% RMS, going to 4th order moments is more than twice as 
efficient \cite{2001PhDT........21S,2002JCoPh.179...27D}. 
Not only does the flop/byte ratio increase with order, but also the ratio of FMA 
(fused multiply add) operations to regular multiply/add, and the number of those compared to 
the one required $1/\sqrt{|{\bf r}|^2}$ increases substantially. The local expansion of the potential 
about the sink's center of mass is actually done to 5th order, 
but we do not store this in the tree, since it is 
sufficient to keep it as a local variable accumulating the CC and CP interactions as we walk 
the tree. We use single precision in calculating interactions, but all components 
are accumulated in double precision so we can achieve force errors of around $10^{-5}\%$, 
well below what is needed for these simulations. To implement periodic boundary conditions, 
PKDGRAV3 uses a 5th order multipole approximation of the Ewald summation potential
\cite{2001PhDT........21S,1991ApJS...75..231H,1997ASPC..123..169K}.
This requires virtually no data movement and is ideally suited to GPU acceleration, but 
these calculations must all be done in double precision.
Our mixed precision approach serves both to reduce memory usage as well as maximizing 
the benefit from AVX/SSE as well as GPU floating point hardware. 

\subsection{Multiple Time Stepping with Dual Trees}

\begin{figure}
  \begin{center}
  \begin{tikzpicture}[>=stealth,thick,scale=0.98]
    \begin{scope}[color=red]
      \foreach \x in {-4,-3,-2,-1,0,1,2,3}{
         \draw[<-] (\x,0.5) -- +(0,0.25);
         \draw (\x,0.5) arc (90:0:0.5cm);
	 }
      \foreach \x in {-3.5,-2.5,-1.5,-0.5,0.5,1.5,2.5,3.5}
         \draw (\x,0) arc (180:90:0.5cm);

      \foreach \x in {-4,-2,0,2}{
         \draw[<-] (\x,1) -- +(0,0.25);
         \draw (\x,1) arc (90:0:1cm);
	 }
      \foreach \x in {-3,-1,1,3}
         \draw (\x,0) arc (180:90:1cm);

      \draw (-4.3,1.5) node [anchor=east,rotate=90] {Very Active};
    \end{scope}

      \foreach \x in {-4,0}{
	\begin{scope}[color=blue]
	\draw (\x,2) -- ++(2,0) node [anchor=south]{Inactive Tree};
	\end{scope}
         \draw[<-] (\x,2) -- +(0,0.25) node [anchor=south]{Kick};
         \draw (\x,2) arc (90:0:2cm);
	 }
      \foreach \x in {-2,2} {
	\begin{scope}[color=blue,->]
	 \draw (\x,2) -- +(-1,-1);
	 \draw (\x,2) -- +(0,-0.5);
	 \draw (\x,2) -- +(1,-1);
	 \end{scope}
         \draw (\x,0) arc (180:90:2cm);
	 }
      \foreach \x in {-4}{
         \draw[<-] (\x,4) -- +(0,0.25) node [anchor=south]{Kick};
         \draw (\x,4) arc (90:0:4cm);
	 }
      \foreach \x in {0}
         \draw (\x,0) arc (180:90:4cm);

      \draw[->] (-4,-0.1) -- (0,-0.1) node [anchor=north]{Time} -- (4,-0.1); 
  \end{tikzpicture}
  \end{center}
\caption{The kick-drift-kick multi-stepping ``umbrella'' diagram with the use of dual trees 
over a single base time-step. Each level of arcs represents one rung and domain decomposition
is allowed to move particles between threads {\em only} at the apex of the black arcs. 
At these points a single tree is built to for all particles in the usual way. 
Next an inactive (or fixed) tree is built halfway through the black interval and used to 
calculate force contributions to the remaining red time-steps in a time symmetric way as shown 
in blue. The red, very active, subtree is all that is built on the shorter very active time-steps
where both trees are walked to obtain the combined force.}
\label{fig:KDK}
\end{figure}
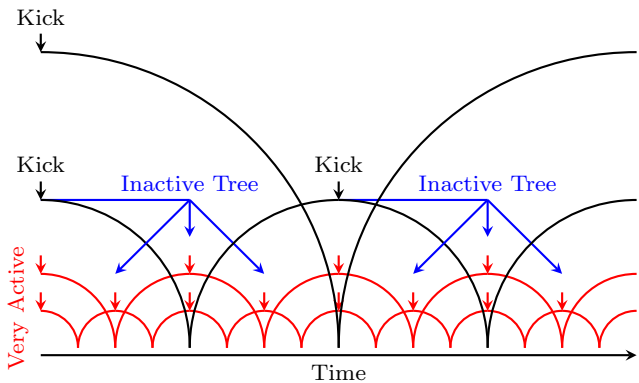

Cosmological simulations span enormous ranges in density, from 
very underdense voids, to the centers of dark matter halos that can have densities 
of 5 orders of magnitude above the mean. This in turn implies that a huge range in
dynamical time-scales exist within the simulation. Calculating gravity on all particles 
at every smallest time-step, while simple from the parallel computing stand-point is 
very wasteful if the the goal is fast time-to-solution for such simulations.
PKDGRAV3 uses individual time-steps per particle, but restricted to being $2^{-l}$ 
times a certain {\em base time-step}, where $l$ is the {\em rung} to which a particle 
belongs. All simulations presented here use 100 equal base time-steps in proper time
to evolve the simulated universes to the present, but many more time-steps are chosen
for dynamically active areas of the simulation automatically.
We use a hierarchical kick-drift-kick leap-frog scheme shown in figure~\ref{fig:KDK},
where the arrows indicate the force calculations that are applied to advance the 
velocities. Only the sink cells that contain particles belonging to rung 
$l$ and higher need to be walked since kicks at higher rungs align in the diagram 
(we call these the {\em active} particles). 
We also need a time-step criterion to decide on which time
scale a particle is evolving. The traditional one used in cosmology simulations is 
based on the particle's softening and the magnitude of its acceleration by 
$\Delta T_i = 0.2 \sqrt{\epsilon / |{\bf a}_i|}$. It has been shown that the power 
spectrum \cite{Schneider:2015we} and mass functions of dark matter halos \cite{2013MNRAS.431.1866R} 
converge using this time-stepping criterion.
Given the distribution of particles in the rungs of a cosmological simulation, the
potential speed-up that is theoretically possible is very large. However, due to the 
ever greater load imbalance, the decreasing flops/byte and the increase in
the relative cost of overheads as the percentage of active particles decreases
makes the speed-ups due to multi-stepping less dramatic, but still often a factor 
of 5x over much of the simulation. We discuss a novel method of reducing the most 
significant overhead, namely the tree build time, by building a second smaller 
tree only for very active particles. 

With any multi-stepping code, there will be rungs with very little gravity work to do
since only a small percentage of the particles are active. 
Nevertheless, the tree must still be built, walked,
and the forces evaluated. The time needed
for the force evaluations reaches a trivial stage while building a full tree
still takes the same amount of time. As the number of tree builds scales as $2^l$,
the tree build cost quickly starts to dominate. We build a single second
very active tree when the number of particles on a rung drop below a certain threshold 
(5\% seems to be a good value)\footnote{The dual
trees are only constructed if there are at least two rungs below the fixed rung,
otherwise there is no performance benefit.}.
The inactive particles are drifted half-way along their trajectory and a fixed tree built
as shown in figure~\ref{fig:KDK}.
Subsequently, only an active tree is built until it is time to kick the fixed particles
at which point they are drifted through the remaining half of their trajectory.
It is very important to construct the second tree by traversing the fixed tree and 
using the same geometric structure. This assures that cells in the very active tree are 
approximately the same size as cells in the fixed tree in a given region of space 
(somewhat similar to the construction of {\em graded} trees in AMR codes). 
Not doing this sometimes results in an unreasonably high number of interacting particles.


\subsection{GPU Acceleration}
While other codes\cite{2012ASPC..453..325B} have attempted to use the GPU for tree related operations,
we made the deliberate decision to split the work between the CPU and GPU in
a manner that compliments their strengths. Walking a tree is geometrically 
complex, exhibits branch divergence, and requires accessing tree nodes on 
remote processors. Conversely, evaluating interactions
and multipoles is ideal work for the GPU.
The GPU work consists of PP interactions, PC interaction
and the periodic boundary condition evaluation (Ewald). PKDGRAV3 monitors the 
flop/byte ratio of interaction lists as they are generated and in the rare case 
that this falls below an optimal threshold then the work is instead issued 
directly to the CPU. This allows the GPU to concentrate
on work packages that can keep utilization high resulting in a lower overall
run-time.
The operations are fully asynchronous allowing almost perfect overlap of compute and
communication with the GPU.

\subsection{Memory}
With the use of FMM, multiple time-steps and GPU acceleration the major limiting factor for 
these simulations is the amount of available memory on each node. PKDGRAV3 has been 
developed to minimize memory usage per particle (see below) and allow the maximal use 
of the available memory for particles. This includes: 1- by-passing Linux file I/O 
and instead using direct I/O to have complete control of file buffering,
2- making memory balancing the primary goal of domain decomposition, 3- reducing the memory
usage by the tree, 4- partitioning memory very carefully on a node and in most cases 
preallocating it.
Careful consideration is also given to the memory usage of the many 
analysis tasks that are performed during the run including group finding, light cone 
generation as well as the storage required to generate the initial condition at the 
beginning of the simulation.\footnote{PKDGRAV3 uses the 2LPT method which requires 13 FFT
operations and with some juggling can be done with 36 bytes per particle.}
Minimizing the memory use per particle has the nice side benefit of increasing performance
in the tree building and tree walking phases of the code that are strongly affected by
the efficiency of transferring to and from memory.

\begin{table}[h!]
  \centering
  \caption{Memory requirements per particle.}
  \label{tab:memory}
  \begin{tabular}{cccc}
    \toprule
    Persistent & Ephemeral & Tree & Buffers\\
    28 bytes & 0-8 bytes & 25 bytes & $\sim$ 5 bytes\\
    \midrule
    \multicolumn{4}{p{3in}}{Buffers are $\mathcal{O}$(125 MB) per thread. 
Here we assume 16 threads with ${\bf 5 \times 10^8}$ particles on a 32 GB node.}\\
    \bottomrule
  \end{tabular}
\end{table}
Storage for particles is divided into two regions; a ``persistent'' area containing properties
that must persist between steps, and ``ephemeral'' storage used for
certain algorithms, for example group finding, where the intermediate data can be forgotten 
when the calculation ends.
In the persistent storage, we identified {\em position}, {\em velocity}, {\em group~id},
and current {\em rung}. Velocities can be stored as single-precision float
values without affecting the results. Positions are trickier. It is necessary to resolve well
below the softening scale which in our case is one part in a million\footnote{Grid size of
$1/20000$ $\times$ softening scale of $1/50$}. We would like to achieve a resolution of perhaps
a hundredth of the softening length which would require of order 27 bits of precision,
greater than that provided by single precision.
We convert double precision float
values between integer coordinates which provides 32 bits\footnote{actually slightly less
as the representable box must be slightly larger than the simulation volume} of precision
which is more than sufficient. We have checked that this simple particle compression 
scheme does not affect their trajectories in any significant way for cosmological 
$N$-body simulations.
The ephemeral storage can vary between zero bytes (when no analysis is required), to
4~bytes if power spectra or group finding is needed up to 8 bytes for other algorithms.
Future analysis may require more memory in which case the ephemeral area would increase.
As a special case, it is possible to use part of the tree memory for algorithms when a
tree is not required (when generating initial conditions for example).
We also need a small amount of memory for explicit communication buffers as well as room
for the tree (which tends to grow as structure forms). All told, the simulation can be run
with approximately 62 bytes per particle as summarized in table~\ref{tab:memory}. 
A simulation of 2 trillion particles can be easily run on Piz Daint (which has 169 TB of memory) 
while an 8 trillion particle simulation can be run on Titan (which has 584 TB).

\section{Performance Results}

\pgfplotstableread{euclid.dat}\euclid

At the time this paper was written, Titan (Oak Ridge National Labs, USA) was the second fastest supercomputer 
in the world with a measured LINPACK performance of 17.59 Pflops and was used for 
most of the performance benchmarks reported here.
It is a Cray XE7 system with 18'688 compute nodes and a Gemini 3-D Torus network. 
Piz Daint (Swiss National Supercomputing Center), a Cray XC30 with 5'272 compute nodes connected via 
the Aries Dragonfly (multilevel all-to-all) network is currently the 7th fastest computer
in the world and is being used for the $2 \times 10^{12}$ particle production run, 
upon which the benchmarks are based (the same mass resolution).
The 282 node Cray XE6, T\"odi (Swiss National Supercomputing Center), is useful for development 
and testing of large scale applications for Titan, being a much smaller instance 
of this system.
The individual nodes of these three machines are similar, each having 32GB of main 
memory a single CPU as well as an nVidia K20X GPU accelerator. 
Titan and T\"odi use the AMD Opteron models 6274 and 6272 with a clock speed 
of 2.2 and 2.1~GHz respectively while Piz Daint uses an Intel Xeon E5-2670 with a 
variable clock speed ranging from 2.6 GHz up to 3.3 GHz 
(3.0 GHz with all cores active).
Titan has the largest total system memory of 584 TB which allows for a production 
simulation with PKDGRAV3 of $8 \times 10^{12}$ particles with a time-to-solution 
of 67 hours. The detailed benchmark and scaling results presented below will establish that
such a high resolution simulation is indeed possible within this projected time.

All of these machines have multiple CPU cores on each node, and the trend is for
this number to increase. PKDGRAV3 employs a ``hybrid'' pthreads/MPI model with a single
MPI thread per node, and threads on the same node exchange data using shared 
memory.
While the dedicated MPI thread is only 25\% utilized, not allowing it to participate
in the gravity calculation has the effect of dramatically reducing message latency and 
increases overall performance.

\subsection{Timing Measurements}
In the following sections, timing information is collected through the use of timers
in the code. The run-time is divided into four phases -- load balancing, tree construction,
force evaluations, and analysis. The first three phases are carefully timed and included in
these results. The fourth, analysis, is not included as it can vary significantly depending
on which analysis needs to be performed. 
If more sophisticated analysis ``instruments'' (by which we mean further software to 
perform on-the-fly analysis) were to be attached to PKDGRAV3 then the 
time would increase from the roughly 25\% for our current production simulations.

We also use the high-resolution on-chip timers to measure sub-phases, in particular we
are able to distinguish how much time is spend calculating forces, how much time is spent
waiting for communication requests to complete, and how much time is wasted at the end of a
step because of load imbalance. We discuss the later two only cursorily as they have a
nearly insignificant effect on time-to-solution as shown in figure~\ref{fig:2T}.
The timings for analysis include the necessary I/O; indeed this can easily be seen in the figure where the analysis time suddenly increases as the ``particle light-cone'' begins.
Raw particle output is written to disk only when checkpointing which takes
takes 30~minutes per checkpoint for the two-trillion particle simulation run on Piz Daint.
This accounts for a roughly 5\% cost increase depending on how frequently checkpoints are written.
Initial conditions are also generated by PKDGRAV3 in memory at the start of the simulation,
a procedure which takes approximately 5~minutes.

\subsection{Simulation Accuracy}

\begin{figure}
\centering
\begin{tikzpicture}     
    \begin{axis}[
      xlabel={Step (of 100)},
      ylabel={Time (Node Hours)},
      reverse legend,
      legend columns=1,
      legend style={draw=none,font=\small},
      legend cell align=right,
      legend plot pos=right,
      stack plots=y,
      ymajorgrids=true,yminorgrids=true,grid style=dotted,
      minor y tick num={1},
      area style,
      enlarge x limits=false,
      xmin = 1,
      ymin = 0,
      ymax=6500
      ]
      \addplot[draw=red,fill=red] table [x expr=\thisrowno{0}+1, color=red, y expr=\thisrowno{6}*\thisrowno{20}/7/3600.] {\euclid} \closedcycle;
      \addlegendentry{Force Evaluation};
      \addplot[draw=yellow,fill=yellow] table [x expr=\thisrowno{0}+1, y expr=\thisrowno{7}*\thisrowno{20}/7/3600.] {\euclid} \closedcycle;
      \addlegendentry{Cache Wait};
      \addplot[draw=black,fill=black] table [x expr=\thisrowno{0}+1, y expr=(\thisrowno{3}-\thisrowno{6}-\thisrowno{7})*\thisrowno{20}/7/3600] {\euclid} \closedcycle;
      \addlegendentry{Work Imbalance};
      \addplot[draw=blue,fill=blue] table [x expr=\thisrowno{0}+1, y expr=\thisrowno{1}*\thisrowno{20}/7/3600] {\euclid} \closedcycle;
      \addlegendentry{Load Balancing};
      \addplot[draw=green,fill=green] table [x expr=\thisrowno{0}+1, y expr=\thisrowno{2}*\thisrowno{20}/7/3600] {\euclid} \closedcycle;
      \addlegendentry{Tree Build};
      \addplot[draw=magenta!70,fill=magenta!70] table [x expr=\thisrowno{0}+1, y expr=\thisrowno{4}*\thisrowno{20}/7/3600] {\euclid} \closedcycle;
      \addlegendentry{Analysis};
    \end{axis}
  \node [anchor=west] (note_A) at (1.7,4.8) {\Large\bf A};
  \draw [-stealth, line width=4pt] (note_A) -- ++(1.0,0.0);
\end{tikzpicture}
\caption{Distribution of run-time between various phases of the calculation.
The red, yellow and black regions are force calculation, the blue region is
for balancing the work, the green region is tree build and the magenta region
is on-the-fly analysis.
The feature indicated by {\bf A} is described in Appendix~\ref{appendix:challenges}.}
\label{fig:2T}
\end{figure}
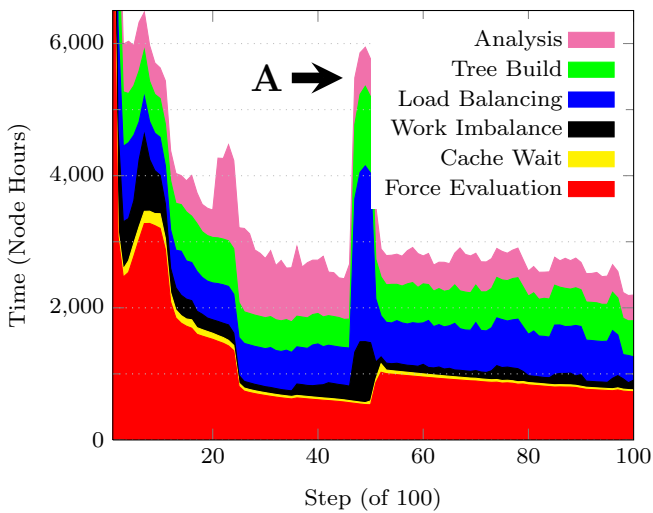

While it is possible to speed-up the simulations by relaxing the accuracy requirements,
taking either fewer time-steps or increasing $\theta$, thereby reducing the force 
accuracy, we emphasize here that we do not do this in any of the benchmarks. We 
run all benchmarks with the {\em same} run parameters that we are using for our 
$2 \times 10^{12}$ particle production simulation which will serve as the first 
reference simulation for the Euclid mission.
At very early times ($z>20$), when the Universe is very homogeneous, the 
forces from opposing directions very nearly cancel and a tree code must use a 
stricter opening criterion in order to attain the same accuracy in the force.
Additionally, small errors in the initial non-linear growth of these first 
structures amplify during the further evolution and can lead to errors greater 
than 1\% in the power spectrum by the end of the simulation if the force accuracy 
and time-stepping is not conservative enough.
We set $\theta=0.40$ for $z > 20$ (to 1\% age of the Universe), 
$\theta=0.55$ for $20 > z > 2$ (to about 20\% age of the Universe), and $\theta=0.70$ 
for the remaining 80\% of the evolution. We note that these quoted $\theta$ values 
apply for the $5^{\rm th}$ order expansion used in PKDGRAV3 and result in much more 
accurate forces than in the traditional quadrupole based BH codes. These transitions
in the force accuracy and cost per step can clearly be seen in figure~\ref{fig:2T}.

The particle mass remained fixed at $10^9$ solar masses for all benchmarks as 
previously mentioned.
This is small enough to converge on the power spectrum to 1\% and to resolve objects 
down to the needed scale to produce so called mock galaxy catalogues \cite{Fosalba:2014di} for 
Euclid, weak lensing maps and statistics for galaxy clusters.
It should be strongly emphasized that the smaller the mass scale that
is simulated, the {\bf harder} the simulation becomes, or comparing simulations of the same $N$, 
the one with the {\bf smaller} box size is the more challenging.
While PKDGRAV3 is independent of the degree of clustering in the force calculation, 
the peak densities within a simulation of smaller particle mass are higher and therefore the 
number of time-steps needed increases. We find that for PKDGRAV3 decreasing the box size
by a factor of two while keeping the same number of particles results in an approximately 50\% longer runtime.

In figure~\ref{fig:2T} we show the actual time spent in different tasks 
integrated over each base time-step for our completed $2 \times 10^{12}$ particle
production run on Piz Daint. Force calculation (in red) dominated the early
time-steps, while later there is a near equal balance between it, 
tree building (green), domain decomposition (blue) and all of the on-the-fly
analysis (magenta).
The yellow/black contribution shows time spent waiting either because the
work is not completely balanced (black), or because of communication delays (yellow).

It used to be the case that analysis was performed by post-processing the results, 
but with the ever increasing simulation sizes writing raw simulation output to 
the disk is no longer feasible, since this would vastly dominate the 
time-to-solution.
The spike in the magenta analysis time at around step 20, for example, is a result of
particle ``light cone'' analysis kicking in. 
Our friends-of-friends group finder, and the analysis on the resulting 
dark matter halos that are found by it, were also completely rewritten to be 
competitive with the other tasks (otherwise it would have been the dominating
task at this scale). It is interesting to see that such analysis tasks must 
not be neglected when considered fast time-to-solution, since even when 
highly optimized, they contribute significantly to the total run time.

While tree building and domain decomposition times remain reasonably
constant, gravity calculation changes for two reasons.
As mentioned previously the force accuracy requirement changes (most 
notably at around step 24) when much of the mass is in viralized dark 
matter halos. The second reason is that the time-step also scales with 
the mean density of the Universe ($\Delta T \propto 1/\sqrt{\rho}$) 
which is decreasing very rapidly early on. This means that at the 
beginning of the simulation there are a lot of particles at very 
small time-step rungs which results in a heftier gravity calculation 
contribution. This never stops so the time per step will
continue to decrease by a modest amount until the very end.
We note again, that this is quite in contrast to what is observed for
BH and P$³$M codes.
The onset of structure formation, which goes in the other direction to 
increase the number of time-steps, can be seen between steps~5 and 10 
when the gravity time increases even though there has been no change in
the force accuracy during this time. Structure formation stabilizes,
in the sense that all density peaks have been established and most of the
mass that can end up in dark matter halos is bound up in them
\footnote{Larger and larger structures continue to form but this does not affect
the time-step hierarchy.}. Finally, the modest cost of tree building
seen here is only possible when using the dual tree method described 
previously. Without this innovation the tree build contribution would 
be 3 times larger.

\subsection{Multi-Stepping and Dual Tree Boost}
Although there were $100$ base steps, PKDGRAV3 uses a multi-stepping scheme where particles
choose their own time-step rung based on the time-step criterion discussed previously.
For the benchmark simulations this results in effectively $5000\pm 10\%$ time steps.
For rungs with very few particles, each step can take a fraction of a second.
While the time for a full gravitational calculation can be in the range of minutes,
the average time per step is of order 50 seconds, including tree build and domain
decomposition (but not including on-the-fly analysis). For simulations of this
type, multi-stepping results in an effective speed-up of between 4x and 5x
when compared to taking single time-steps.

As discussed earlier, the tree building phase can begin to dominate when
multi-stepping. A complete gravity takes of order two minutes, while constructing
the tree takes more like 25 seconds. When multi-stepping, some of the gravity
calculation takes less than a second while the tree building time does not vary.
By constructing a second tree for the very active particles, the tree build time
is reduced to one second for these critical sub-steps. The method results in
an additional 26\% decrease in the overall time-to-solution.

\subsection{GPU Boost}
PKDGRAV3 is already highly optimized for SIMD type instructions,
such as SSE and AVX, and because of mixed-precision (float/double) code, the performance
boost is already a factor of eight for some parts of the calculations. Because
not all calculation are FLOP dominated, for example load balancing and tree construction,
the effective speed-up is more like 3x.
By using the GPU, the situation is dramatically improved.
For the T\"odi simulation shown in figure~\ref{fig:N2500}, a single force evaluation\footnote{At late time when gravity calculations
no longer dominate the run-time; speed-up at earlier times is higher.}
that took 1138 seconds using only the CPU,
takes 119.5 seconds when using the GPU -- a speed-up of 9.5x. A complete
step, including all phases (gravity, tree construction and load balancing),
takes 1629 seconds with the GPU compared to 6507 with the CPU only,
resulting in a 4.0x improvement in the time-to-solution.

Part of the GPU work scheduling involves shunting work to the CPU when appropriate.
If the number of particles is too small (1 or 2), then the CPU will do the work.
If the GPU is too busy, detected when too many work packages are scheduled on
the GPU but not yet complete, then pieces of the interaction list that do not
evenly align with a WARP\footnote{If the interaction list has 655 elements
for example, then 640 would be calculated by the GPU, and 15 by the CPU.}
are done by the CPU instead.
While it is possible to push more work to the GPU, and thus increasing the total
FLOP rate, this comes at the expense of an increased time-to-solution.

\subsection{Scaling}

To perform the very largest simulations, it must be demonstrated that
PKDGRAV3 can efficiently scale up to the task.
Weak scaling was measured by starting with a $1000^3$ simulation ($10^9$ particles) and running it on two nodes
to measure the gravity calculation times. The simulation was then scaled upward by scaling the total number of
particles and the total number of nodes by the same factor.
The simulations run are outlined in table~\ref{tab:weaktitan}. Here we see that the total run time remains
constant as the simulation size is increased, which is expected for an $\mathcal{O}(N)$ method which has low
parallel overheads and good load balance. 
We include a direct comparison with the HACC\cite{2013hpcn.confE...6H,2016NewA...42...49H}
and 2HOT\cite{2014arXiv1407.2600S} codes. 
The weak scaling runs for PKDGRAV3 were all performed with $4.7\times 10^8$ particles per node,
the HACC benchmarks with $0.32\times 10^8$ particles per node,
and the 2HOT simulation with $0.81\times 10^8$ particles per node.
As the weak scaling of these codes is essentially perfect,
the total run-time does not change when using the same
number of particles per node. This is the most relevant scaling for these types of cosmological 
simulations as it is typical to be memory limited due to the desire for high resolution as well as
large volume.
For the same simulation size, $1.0\times 10^{12}$ particles,
the results from HACC, 2HOT and PKDGRAV3 are similar with a science rate
(millions of particles per second per node) of
$1.7$ for HACC\footnote{Private communication},
$1.2$ for 2HOT\footnote{Table 1 of \cite{2014arXiv1407.2600S}},
and $3.8$ for PKDGRAV3. As the HACC and 2HOT benchmarks are not particularly current
we would expect that today improved results could be presented by these authors.
When the total number of particles per node
was kept fixed at $4.7\times 10^8$ as was the case for the weak scaling tests,
an entire simulation would run to completion in 67 hours {\bf regardless of size}.
\begin{table}[ht!]
  \centering
  \caption{Weak Scaling Performance on Titan with $4.7\times 10^8$ particles per node.
The science rate remains constant.}
  \label{tab:weaktitan}
  \begin{tabular}{rrrr|rr}
    \toprule
    & & & & \multicolumn{1}{c}{Science} \\
     Nodes & \multicolumn{1}{c}{$N_p$} & Mpc & Time & \multicolumn{1}{c}{Rate\footnotemark}  \\
    \midrule
       $2$  &  $1.0\times 10^{9}$ &  $250$ & $124.9$ & $4.00\times 10^6$ \\
      $17$  &  $8.0\times 10^{9}$ &  $500$ & $117.4$ & $4.02\times 10^6$ \\
     $136$  &  $6.4\times 10^{10}$ & $1000$ & $117.9$ & $3.98\times 10^6$ \\
    \midrule
     $266$  &  $1.3\times 10^{11}$ & $1250$ & $125.1$ & $3.76\times 10^6$ \\
    $2125$  & $1.0\times 10^{12}$ & $2500$ & $124.0$ & $3.79\times 10^6$ \\
    $7172$  & $3.4\times 10^{12}$ & $3750$ & $123.2$ & $3.82\times 10^6$ \\
   $11390$  & $5.4\times 10^{12}$ & $4375$ & $126.6$ & $3.72\times 10^6$ \\
    \midrule
   $18000$  & $8.0\times 10^{12}$ & $5000$ & $120.1$ & $3.70\times 10^6$ \\
    \bottomrule
  \end{tabular}
\end{table}
\footnotetext{in particles per second per node}

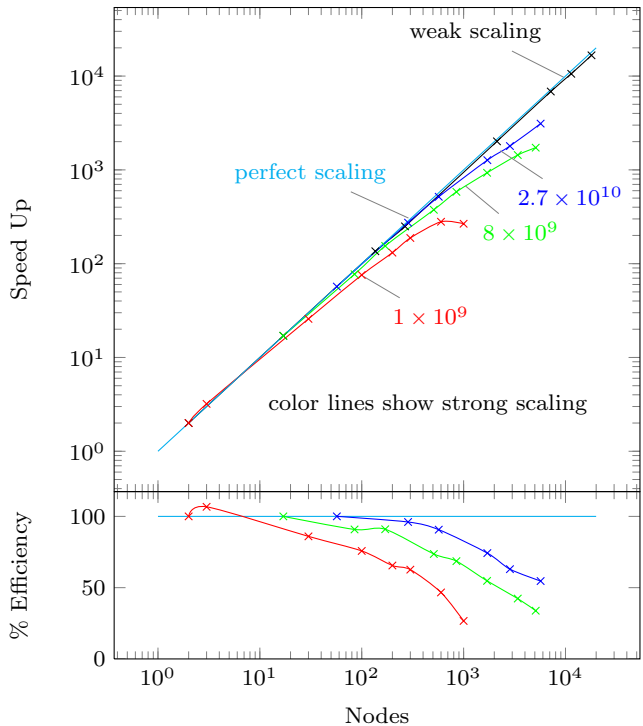
\begin{figure}[ht]
\pgfplotstableread{weak1000.dat}\weak
\pgfplotstablevertcat{\weak}{weak2500.dat}
\pgfplotstablegetelem{0}{0}\of{\weak}
\pgfmathsetmacro{\N}{\pgfplotsretval}
\pgfplotstablegetelem{0}{1}\of{\weak}
\pgfmathsetmacro{\Np}{\pgfplotsretval}
\pgfplotstablegetelem{0}{2}\of{\weak}
\pgfmathsetmacro{\WALL}{\pgfplotsretval}

\pgfplotstableread{strong2.dat}\strongA
\pgfplotstablegetelem{0}{0}\of{\strongA}
\pgfmathsetmacro{\NpA}{\pgfplotsretval}
\pgfplotstablegetelem{0}{1}\of{\strongA}
\pgfmathsetmacro{\WallA}{\pgfplotsretval}

\pgfplotstableread{strong2000.dat}\strongB
\pgfplotstablegetelem{0}{0}\of{\strongB}
\pgfmathsetmacro{\NpB}{\pgfplotsretval}
\pgfplotstablegetelem{0}{1}\of{\strongB}
\pgfmathsetmacro{\WallB}{\pgfplotsretval}

\pgfplotstableread{strong3000.dat}\strongC
\pgfplotstablegetelem{0}{0}\of{\strongC}
\pgfmathsetmacro{\NpC}{\pgfplotsretval}
\pgfplotstablegetelem{0}{1}\of{\strongC}
\pgfmathsetmacro{\WallC}{\pgfplotsretval}

\begin{tikzpicture}[scale=1.0]
\begin{groupplot}[
    group style={
        group name=my plots,
        group size=1 by 2,
        xlabels at=edge bottom,
        xticklabels at=edge bottom,
        vertical sep=0pt
    },
    width=8.5cm,
    height=10.8cm,
    xlabel={Nodes},
    ylabel={Speed Up}]
\nextgroupplot[width=8.5cm,height=8cm,xmode=log,ymode=log]

\addplot[black,smooth,mark=x]
table[x index=1,
      y expr=\thisrowno{0}^3/\thisrowno{2} / (\N^3/\WALL) * \Np ]
      {\weak}
  node at (axis cs:1e4,0.95e4) [pin={120:weak scaling},inner sep=0pt] {};

\addplot[red,smooth,mark=x]
table[x index=0,
      y expr= \WallA * \NpA / \thisrowno{1} ]
      {\strongA}
  node at (axis cs:1e2,0.75e2) [pin={-45:$1\times 10^9$},inner sep=0pt] {};

\addplot[green,smooth,mark=x]
table[x index=0,
      y expr= \WallB * \NpB / \thisrowno{1} ]
      {\strongB}
  node at (axis cs:1e3,0.7e3) [pin={-70:$8\times 10^9$},inner sep=0pt] {};

\addplot[blue,smooth,mark=x]
table[x index=0,
      y expr= \WallC * \NpC / \thisrowno{1} ]
      {\strongC}
  node at (axis cs:2.2e3,1.65e3)  [pin={-70:$2.7\times 10^{10}$},inner sep=0pt] {};

\addplot[cyan,no marks][domain=1:2e4]{x}
  node at (axis cs:3e2,3e2) [pin={120:perfect scaling},inner sep=0pt] {};

\node at (axis cs:10,5) [anchor=north west] {color lines show strong scaling};


\nextgroupplot[width=8.5cm,height=3.8cm,xmode=log,ylabel=\% Efficiency,ymin=0]
\addplot[cyan,no marks][domain=1:2e4]{100};
\addplot[red,smooth,mark=x]
table[x index=0,
      y expr= 100 * \WallA * \NpA / \thisrowno{1} / \thisrowno{0} ]
      {\strongA};

\addplot[green,smooth,mark=x]
table[x index=0,
      y expr= 100 * \WallB * \NpB / \thisrowno{1} / \thisrowno{0}]
      {\strongB};

\addplot[blue,smooth,mark=x]
table[x index=0,
      y expr= 100 * \WallC * \NpC / \thisrowno{1} / \thisrowno{0}]
      {\strongC};

\end{groupplot}
\end{tikzpicture}
\caption{Weak and Strong Scaling. Perfect $\mathcal{O}(N)$ scaling should follow the slope of the ``perfect scaling'' line.
PKDGRAV3 exhibits perfect weak scaling and excellent strong scaling out to 300 times the number of nodes.
This translates into node memory usage starting at 30 GB and scaling to 0.1 GB.}\label{fig:weak}
\end{figure}

To measure strong scaling, we start with a series of simulations with $1000^3$, $2000^3$ and $3000^3$
particles ($10^9$, $8\times 10^9$ and $2.7\times 10^{10}$) and run them on the smallest number
nodes where they will fit (so $4.7\times 10^8$ particles per node). The number of nodes is then
incrementally increased. As shown in figure~\ref{fig:weak}, PKDGRAV3 shows excellent strong
scaling up to a factor of several hundred. This allows us to reduce the wall-clock time of
simulations by up to a factor of a hundred or more by simply increasing the number of nodes.
Recall that when using the most particles possible per node and hence the maximum wall clock
time, a simulation will take approximately 67 hours. Using $10$ times as many nodes results in
only a $25\%$ penalty meaning a simulation would take less than $10$ hours. Using $100$ times
as many nodes carries a $70\%$ penalty, meaning a simulation would take slightly longer than
an hour.

\subsection{Raw Performance}
With PKDGRAV3, a great deal of effort has gone into algorithmic improvements
to try to avoid, wherever possible, doing unnecessary work. This has the
effect of greatly complicating the data structures making it more
difficult to achieve high raw flop counts. Nevertheless, for a code to
achieve high performance, the raw performance must be at least competitive.

To determine the number of floating point operations used, the AVX version
of the code was examined to determine how many floating point instructions
were required for each phase of the calculations. Most operations, including
addition, subtraction and multiplication count as a single flop.
The reciprocal square root is scored as seven flops while a division is scored
as 35 flops.  The totals for each phase are shown in table~\ref{tab:flopcounts}.
In addition, floating point operations were divided into single
and double precision, and totaled separately for the CPU and GPU.
\begin{table}[ht!]
  \centering
  \caption{flop counts by phase}
  \label{tab:flopcounts}
  \begin{tabular}{lrrrr}
    \toprule
    Phase & $+-\times$ & $\surd$ & $\div$ & FLOPs \\
    \midrule
    Particle/Particle &  46 & 1 &   & 53\\
    Particle/Cell     & 208 & 1 &   & 215\\
    Cell/Particle     & 206 & 1 &   & 213\\
    Cell/Cell         & 472 & 1 &   & 479\\
    Ewald iteration   & 433 & 1 & 2 & 510\\
    Opening criteria  &  97 &   &   & 97\\
    \bottomrule
  \end{tabular}
\end{table}

In table~\ref{tab:peaktitan} we show the peak performance achieved for various
simulation sizes where the number of particles is optimized to fill a node.
We also show the wall-clock time required to calculate the forces for
a single particle.

\begin{table}[ht!]
  \centering
  \caption{Performance on Titan.
    Total measured TFlops as well as the wall-clock time to calculate the forces for a single particle. }
  \label{tab:peaktitan}
  \begin{tabular}{rrrrr}
    \toprule
    Nodes & $N_p$ & Mpc & TFlops & Time / Particle \\
    \midrule
       $2$  & $1.0\times 10^{9}$ &  $250$ &    $1.2$ & $125$ $ns$  \\
      $17$  & $8.0\times 10^{9}$ &  $500$ &   $10.3$ & $14.7$ $ns$ \\
     $136$  & $6.4\times 10^{10}$ & $1000$ &   $82.2$ & $1.84$ $ns$ \\
    \midrule
     $266$  & $1.3\times 10^{11}$  & $1250$ &  $152.5$ & $1.00$ $ns$ \\
    $2125$  & $1.0\times 10^{12}$ & $2500$ & $1230.3$ & $0.124$ $ns$ \\
    $7172$  & $3.4\times 10^{12}$ & $3750$ & $4130.9$ & $0.0365$ $ns$ \\
   $11390$  & $5.4\times 10^{12}$  & $4375$ & $6339.2$ & $0.0236$ $ns$ \\
    \midrule
   $18000$  & $8.0\times 10^{12}$ & $5000$ & $10096.2$ & $0.0150$ $ns$ \\
    \bottomrule
  \end{tabular}
\end{table}

While PKDGRAV3 does use mixed precision float code, the measured
10 Pflops compares quite well with the 17.59 measured LINPACK performance.

\subsection{Time to Solution}

\begin{figure}[ht]
\begin{tikzpicture}[scale=1.0]
\begin{semilogyaxis}[
        xlabel={Step},
        ylabel={Node Minutes}]
\addplot[red,smooth,mark=o]
table[x index=0,y expr=(\thisrowno{19}) * 320 / 60]
  {N2500.dat}
  node [left] at (axis cs:105,1.1e2) {$1.5\times 10^{10}$};

\addplot[smooth,dashed]
table[x index=0,y expr=((5000/2500)^3*\thisrowno{19}) * 320 / 60]
  {N2500.dat}
  node [left] at (axis cs:105,8e2) {$1.25\times 10^{11}$};
\addplot[scatter,only marks]
coordinates{
(50,125.050125*266/60)
(20,207.420066*266/60)
(0,349.024972*266/60)
};

\addplot[smooth,dashed]
table[x index=0,y expr=((10000/2500)^3*\thisrowno{19}) * 320 / 60]
  {N2500.dat}
  node [left] at (axis cs:105,6e3)  {$1\times 10^{12}$};
\addplot[scatter, only marks]
coordinates{
(50,123.981763*2125/60)
(20,203.239562*2125/60)
(0, 362.132831*2125/60)
};

\addplot[smooth,dashed,restrict expr to domain={x}{0:76}]
table[x index=0,y expr=((15000/2500)^3*\thisrowno{19}) * 320 / 60]
  {N2500.dat}
  node [left] at (axis cs:105,1.4e4) {$3.4\times 10^{12}$};
\addplot[scatter, only marks]
coordinates{
(50,123.190040*7172/60)
};

\addplot[smooth,dashed,restrict expr to domain={x}{0:76}]
table[x index=0,y expr=((17500/2500)^3*\thisrowno{19}) * 320 / 60]
  {N2500.dat}
  node [left] at (axis cs:105,2.4e4) {$5.4\times 10^{12}$};
\addplot[scatter, only marks]
coordinates{
(50,126.614430*11389/60)
};

\addplot[smooth,dashed]
table[x index=0,y expr=((20000/2500)^3*\thisrowno{19}) * 320 / 60]
  {N2500.dat}
  node [left] at (axis cs:105,5e4) {$8\times 10^{12}$};
\addplot[scatter, only marks]
coordinates{
(0,416.755539*18000/60)
(20,216.636079*18000/60)
(50,120.089825*18000/60)
};

\end{semilogyaxis}
\end{tikzpicture}
\caption{Gravity time per step. Circles are measurements while dashed lines are predictions.}\label{fig:N2500grav}
\end{figure}
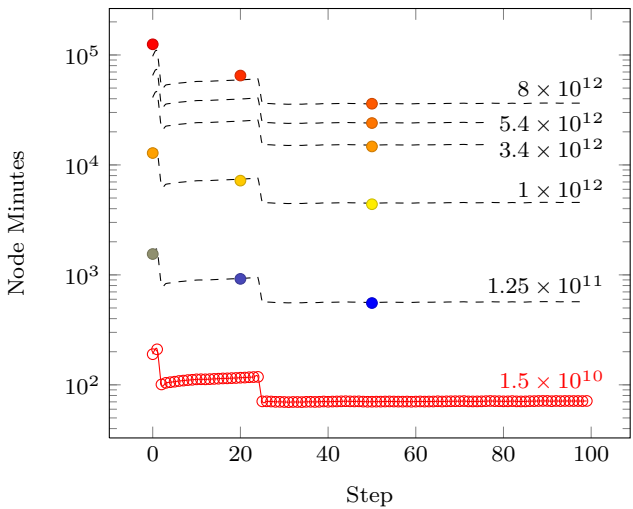

To measure time-to-solution, we start by running a complete simulation at a lower resolution.
Because of the physical processes involved, the timings for each step can be
roughly broken into three distinct phases corresponding to different integration
accuracy domains. In figure~\ref{fig:N2500grav} we show the timings for gravity calculations in total node hours
during each of the 100 main steps. As PKDGRAV3 is an $\mathcal{O}(N)$ code, these
timings are then scaled linearly by the problem size to estimate how long the force
calculations will take.
The estimates are verified by running the force calculation at sampled points, and
comparing them to the estimates.

This can be seen in figure~\ref{fig:N2500grav}. The hollow circles represent
the measured timing for a force calculation on all particles throughout
a simulation of $2500^3$($1.5\times 10^{10}$) particles. As is clearly
apparent in the figure, the time required perform the gravity calculation is
extremely stable. The three different ``steps'' correspond to the accuracy
requirements (high redshift requires increased accuracy). The timings are given
in node minutes (wall clock time multiplied by the number of nodes).

The dashed lines show predictions for the force evaluations at increasing
resolutions made by scaling the low resolution simulation by the problem size.
Measurements were then taken a several points at each resolution shown by
the solid circles. The prediction and measurements agree perfectly.

\pgfplotstableread{N2500.dat}\Nbig
\pgfplotstableread{W4650.dat}\Wsecond

\begin{figure}[ht]
\begin{tikzpicture}[scale=1.0]
\begin{semilogyaxis}[
        xlabel={Step},
        ylabel={Cumulative Node Hours}]
\addplot[blue,only marks,mark=o]
table[x index=0,y expr=\pgfmathaccuma + ((\thisrowno{3} + \thisrowno{2} + \thisrowno{1}) * \thisrowno{20}) / 7 / 60 / 60] {\euclid}
  node [right] at (axis cs:55,1e5) {$2\times 10^{12}$ (Piz Daint)};

\addplot[blue,only marks,mark=o]
table[x index=0,y expr=\pgfmathaccuma + (\thisrowno{3} + \thisrowno{2} + \thisrowno{1}) * 214 / 60 / 60] {\Wsecond}
  node [right] at (axis cs:55,6e3) {$1\times 10^{11}$ (T\"odi)};

\addplot[red,only marks,mark=o]
table[x index=0,y expr=\pgfmathaccuma + (\thisrowno{3} + \thisrowno{2} + \thisrowno{1}) * 320 / 60 / 60] {\Nbig}
  node [right] at (axis cs:55,1e3) {$1.5\times 10^{10}$ (Titan)};

\addplot[smooth,dashed,very thick]
table[x index=0,y expr=\pgfmathaccuma + (4650/2500)^3*(\thisrowno{3} + \thisrowno{2} + \thisrowno{1}) * 320 / 1.24 / 60 / 60] {\Nbig}
  node [left] at (axis cs:105,2.3e4) {$1\times 10^{11}$};

\addplot[smooth,dashed]
table[x index=0,y expr=\pgfmathaccuma + (12600/2500)^3*(\thisrowno{3} + \thisrowno{2} + \thisrowno{1}) * 320 / 1.24 / 60 / 60] {\Nbig}
  node [left] at (axis cs:105,5e5) {$2\times 10^{12}$};

\addplot[smooth,dashed]
table[x index=0,y expr=\pgfmathaccuma + (20000/2500)^3*(\thisrowno{3} + \thisrowno{2} + \thisrowno{1}) * 320 / 1.24 / 60 / 60] {\Nbig}
  node [left] at (axis cs:105,1.5e6) {$8\times 10^{12}$};
\end{semilogyaxis}
\end{tikzpicture}
\caption{Total runtime in node hours (wall clock $\times$ number of nodes).
The red circles are measurements of a simulation run on Titan with 320 nodes.
The dashed lines are predictions based on both weak and strong scaling.
The much larger T\"odi simulation was run on 214 nodes while the even larger
Daint run used 4900, and later 4000 nodes. All measurements show excellent agreement with the predictions.}%
\label{fig:N2500}
\end{figure}
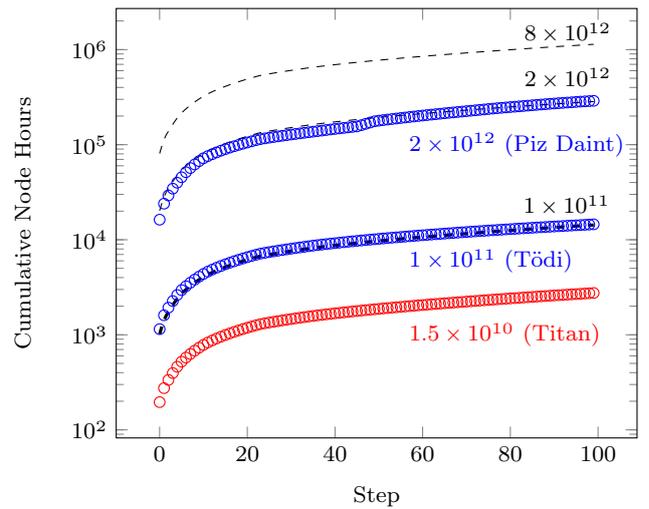

In figure~\ref{fig:N2500}, the cumulative node hours for the reference simulation
is plotted. In order to complete the simulation quickly,
it was run on $320$ nodes, even though it could have fit in as few at $32$.
We make predictions for how long simulations of various sizes, namely $10^{11}$,
$2\times 10^{12}$ and $8\times 10^{12}$ would take based on the weak scaling.
As this is now in the strong scaling regime we further correct the prediction by
assuming that it could be run $24\%$ faster (recall that the penalty for strong
scaling by a factor of $10$ is $24\%$). These predictions are shown as dashed
black lines.

A complete simulation of $10^{11}$ particles was run on T\"odi using $214$ nodes
which corresponds to full memory usage of $4.7\times 10^8$ particles per node.
The measured performance of the simulation shows perfect agreement with the estimate.
A further simulation was run on Piz Daint using $2\times 10^{12}$ ($2$ trillion) particles.
Due to the slight differences in architecture, the simulation actually beats the prediction
by a modest amount. We expect this is due to the slightly better AVX performance on the CPU,
and perhaps to a lesser degree the network.

The end result is that we have high confidence that an $8$ trillion particle simulation
is possible on Titan using $18,600$ nodes, and it will take of order 67 hours with some
additional time for on-the-fly analysis which would vary depending on the exact analysis done.
This also means that a $1$ trillion particle simulation run on Titan using all nodes could be
completed in under $10$ hours.


%
%
%
%
%

\section{Implications}



In order to achieve the results presented here, significant {\em refactoring}
of the code was required. Tracking the progress 
in $N$-body simulations over time, 
a performance doubling time of roughly 1 year is observed. This rate, 
which exceeds Moore's Law can only continue if further efforts are made
to refactor algorithms for new computing hardware. These gains can also be 
pushed forward by {\em co-design}, where computing hardware and algorithmic
developments are considered as a single design process.

The new time-to-solution of these simulations is a game changer as far as 
the way theory is used in cosmological measurements. For the first time 
simulations will not only be used to help understand effects or to make some 
predictions, but will be needed to extract fundamental physical parameters 
from future survey data. They must become part of the data analysis pipelines.

Another implication for the future is that time-to-solution will continue
to decrease as greater computational speed will out-strip any possible
increase in memory size. Our memory footprint is about as low as it is 
possible to go per particle, so that the time-to-solution for these 
simulation can only decrease from this point on.
We expect to run such simulations within 8 hours
or less within the decade. This also means that raw data will never be 
stored and post-processed. Instead data analysis ``instruments'' will be
attached to the code and the simulations will be rerun, perhaps several
times with different ``instrumentation''. This is starting to happen and
is a true paradigm shift in the field of simulations.

\section*{Acknowledgments}
We are indebted to the folks at the Swiss National Supercomputer Center,
specifically Thomas Schulthess, Maria Grazia Giuffreda and Claudio Gheller
for the support and advice, and for resurrecting T\"odi. We would also like
to thank Jack Wells for arranging access to Titan. Parts of this work were
supported by a grant from the Swiss National Supercomputing Center (CSCS)
under project ID S592.
This research used resources of the Oak Ridge Leadership Computing Facility
at the Oak Ridge National Laboratory, which is supported by
the Office of Science of the U.S. Department of Energy
under Contract No. DE-AC05-00OR22725.

\appendix
\section{Computational Challenges}
\label{appendix:challenges}
During Grand Challenge simulations such as this one, there are
inevitably problems encountered, and such was the case here.
In figure~\ref{fig:2T}, the time per step suddenly increases at
step~46 as indicated by the arrow labelled {\bf A}. This was
caused by one of the nodes performing in a substandard way which
resulted in the entire simulation to take twice as long, as the other
nodes were waiting for this node to complete its share of the work.
The exact cause of this problem is not known, and will never be
known, but it was very likely a rogue process that was left running
on the node that stole processing cycles. This problem disappeared
when the simulation was restarted without this node.

The second problem occurred shortly thereafter, around step 50,
and was a result of the increase in efficiency as the simulation
progressed. In figure~\ref{fig:2T} we see that the gravity
calculation time drops dramatically between step 0 and step 20
as structure forms and the effect of the initial condition grid
is no longer relevant allowing the force accuracy to be relaxed.
At some point, the amount of work being shipped to the GPU reaches
a threshold that triggers a not yet understood problem with the
GPU device. When this threshold is reached, the GPU will,
very rarely, accept work but never complete it. By sending work
in a more controlled fashion, this problem is eliminated or vastly
reduced allowing the simulation to run to solution, but with
slightly decreased performance. The cause of this is still
under investigation.

Although these two problems seem dramatic, they had very little
impact on the total run-time as can be seen in figure~\ref{fig:N2500}.
The simulation was on track to slightly
beat the estimate, but the two problems conspired to slightly
increase the total run-time causing it to take almost
exactly the amount of time predicted.


\bibliographystyle{spphysetal}       
\bibliography{pkdgrav3,romain}

\end{document}